\documentclass[twocolumn,twoside,slac_two]{revtex4}
\usepackage{graphicx}
\usepackage{fancyhdr}
\usepackage{graphics}
\usepackage{epstopdf}
\usepackage{textpos}
\begin{document}

\title{\centering High $p_T$ Jet Physics}

\author{
\centering
\includegraphics[scale=0.35]{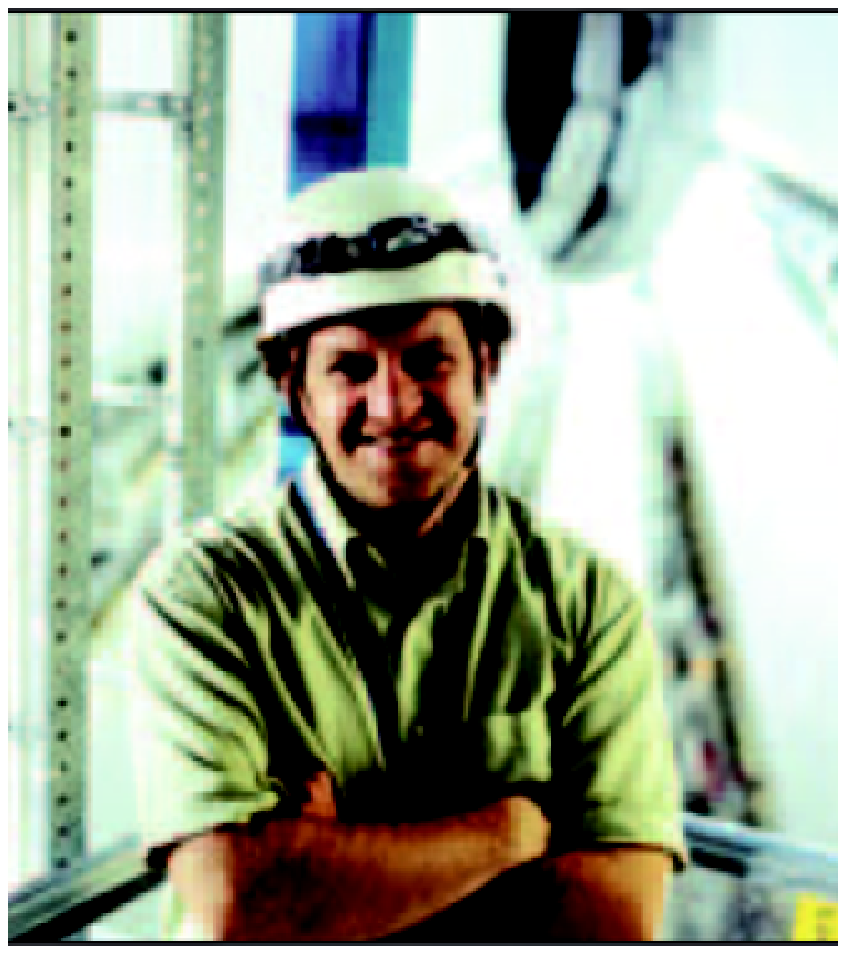} \\
\begin{center}
Richard J. Teuscher
\end{center}}
\affiliation{\centering IPP and University of Toronto, Toronto ON M5S 1A7 Canada}

\begin{abstract}
This report gives a selection of recent jet results from
the LHC and Tevatron, including inclusive jet production, dijets,
and jets produced in association with massive vector bosons.
\end{abstract}

\maketitle
\thispagestyle{fancy}

\section{Introduction}
Jet production is the dominant high transverse-momentum ($p_T$) process at the Tevatron and LHC, and jet measurements are a key
step in searches for physics beyond the Standard Model.
For a jet with energy $E$ and momentum $\vec{p}=(p_x,p_y,p_z)$,  its transverse
momentum is given by $p_T=\sqrt{p_x^2+p_y^2}$, and its rapidity is given by 
$y = \frac{1}{2} \ln{ {\frac{E+pz}{E-pz}}}≈ß$. Jets are defined by jet-finder algorithms \cite{salam}, 
clustered in a two-dimensional space of $y$ and detector azimuthal angle $\phi$,
with a distance parameter defined as
$R = \sqrt{\delta {y}^2 + \delta {\phi}^2 }$. 
 Here we focus on results with the
 anti-kt jet algorithm as used by ATLAS with $(R=0.4,0.6)$ and by CMS with
$(R=0.5, 0.7)$.

Due to the non-compensating nature of typical calorimeters, such that the ratio of
electromagnetic to hadronic response differs from unity, jets must be calibrated
to determine the Jet Energy Scale (JES).
This is usually achieved using
a combination of Monte Carlo and in-situ techniques.
The resulting JES uncertainties are typically of order 1.5\% in the central rapidity
detector regions for CDF/DO, while for ATLAS/CMS they
 range from about 2.5\% for central jets
with ($50 < p_T < 800$) GeV, increasing up to 12\% for forward jets.

In this report, the results from ATLAS and CMS 
comprise up to 48 $pb^{-1}$ of integrated
luminosity delivered by the LHC in 2010.

\section{Inclusive Jets}
\begin{figure*}
\includegraphics[width=80mm]{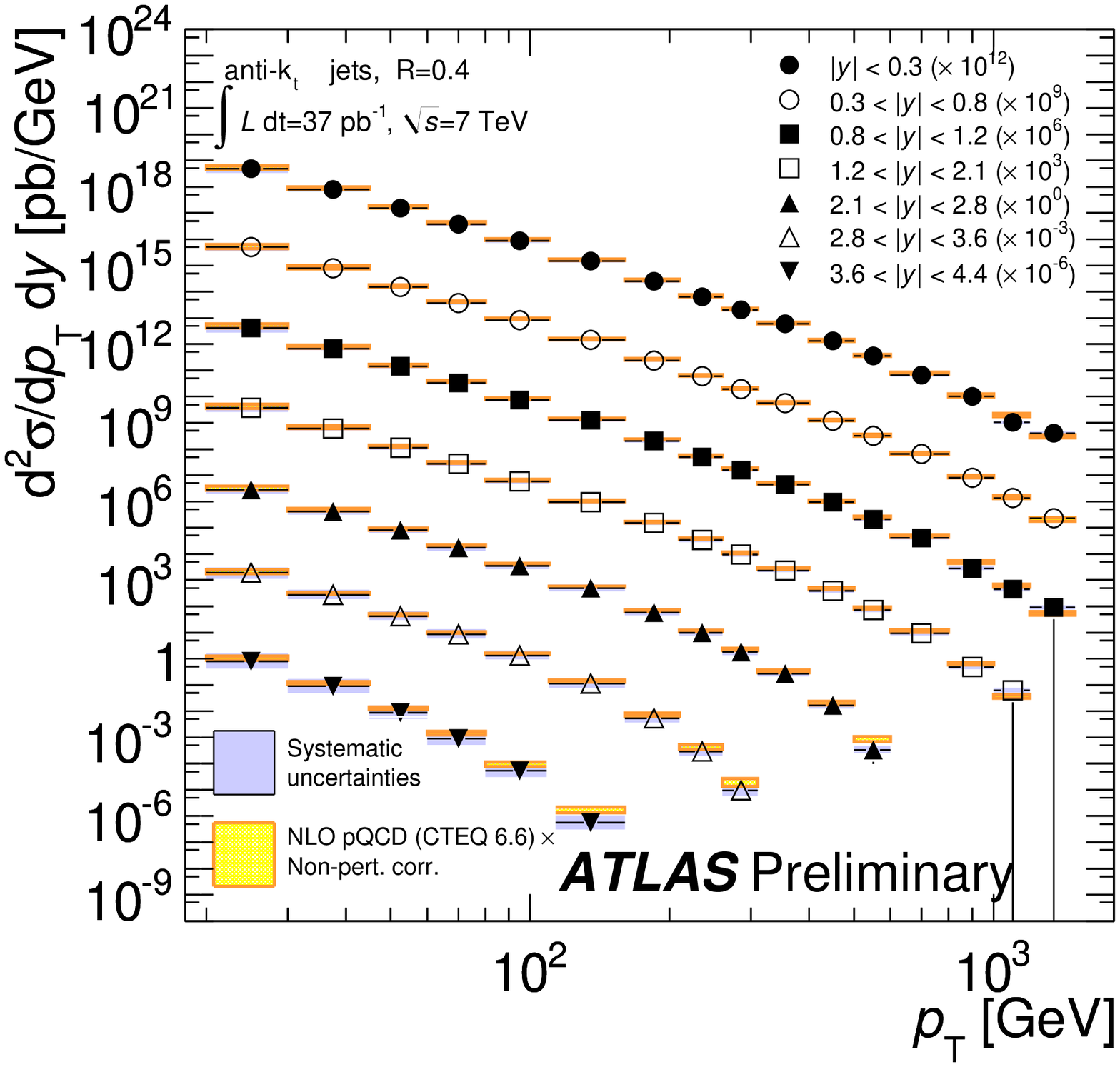}
\includegraphics[width=80mm]{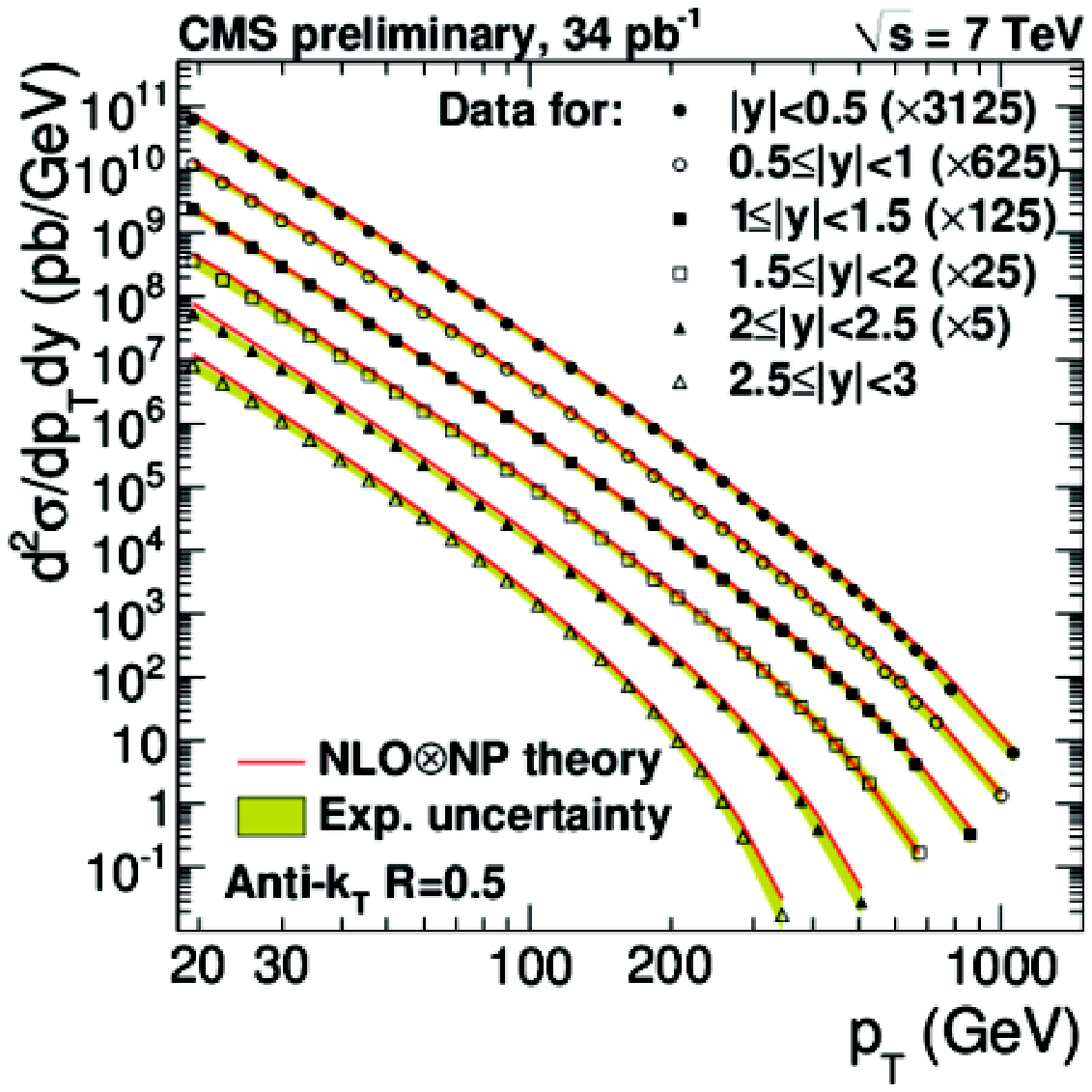}
\caption{Inclusive jet double-differential cross section as a function of jet $p_T$ in different regions of $|y|$ for jets identified using the anti-kt algorithm with $R=0.4$ (ATLAS) and $R=0.5$ (CMS) measured at $\sqrt{s}=7$ TeV. The data are compared to NLO pQCD calculations to which non-perturbative corrections have been applied.  The error bars
indicate the statistical errors on the measurements, and the
dark shaded band indicates the quadratic sum of the experimental systematic
uncertainties.}
\label{f:inclusivejets}
\end{figure*}

Inclusive jet production at the LHC centre-of-mass energies of 
$\sqrt{s}=7$ TeV has been measured by
ATLAS \cite{ATLASinclusive} and CMS \cite{CMSinclusive},
as show in Figure \ref{f:inclusivejets}.  The figure illustrates the inclusive 
jet double-differential cross section as a function of jet $p_T$ in different regions 
of $|y|$ for jets identified using the anti-kt algorithm.  Here the examples
with $R=0.4$ (ATLAS) and $R=0.5$ 
(CMS) are shown. The data are compared to next-to-leading order perturbative QCD (NLO pQCD) calculations
to which non-perturbative corrections have been applied.  The error bars
indicate the statistical errors on the measurements, and the
dark shaded band indicates the quadratic sum of the experimental systematic
uncertainties, dominated by the JES uncertainty.
The measurements cover jet $p_T$ from 20 GeV to 1.5 TeV, rapidities
within $|y|<4.4$, and span six orders of magnitude in cross sections.

\begin{figure*}
\includegraphics[width=80mm]{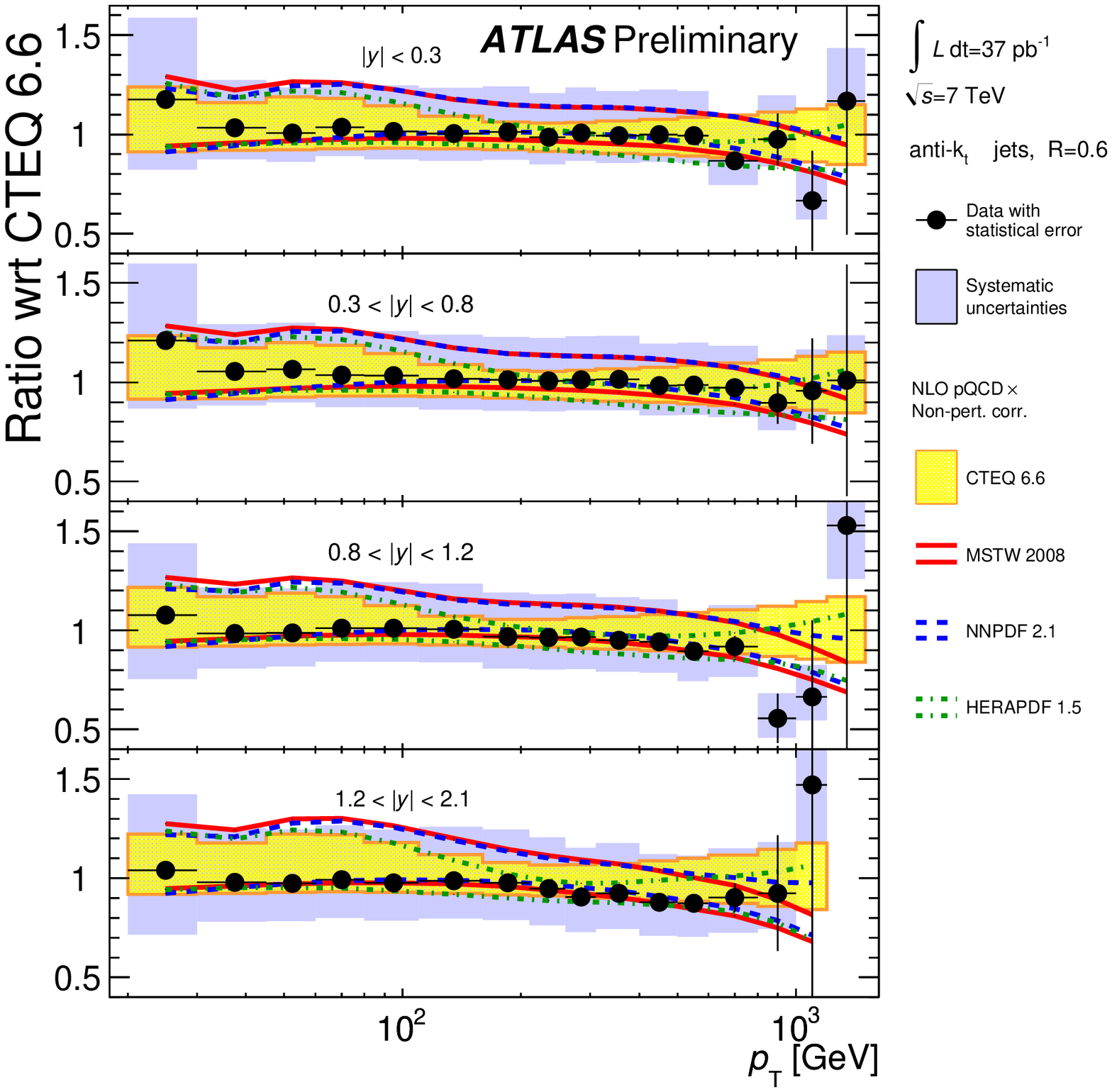}
\includegraphics[width=80mm]{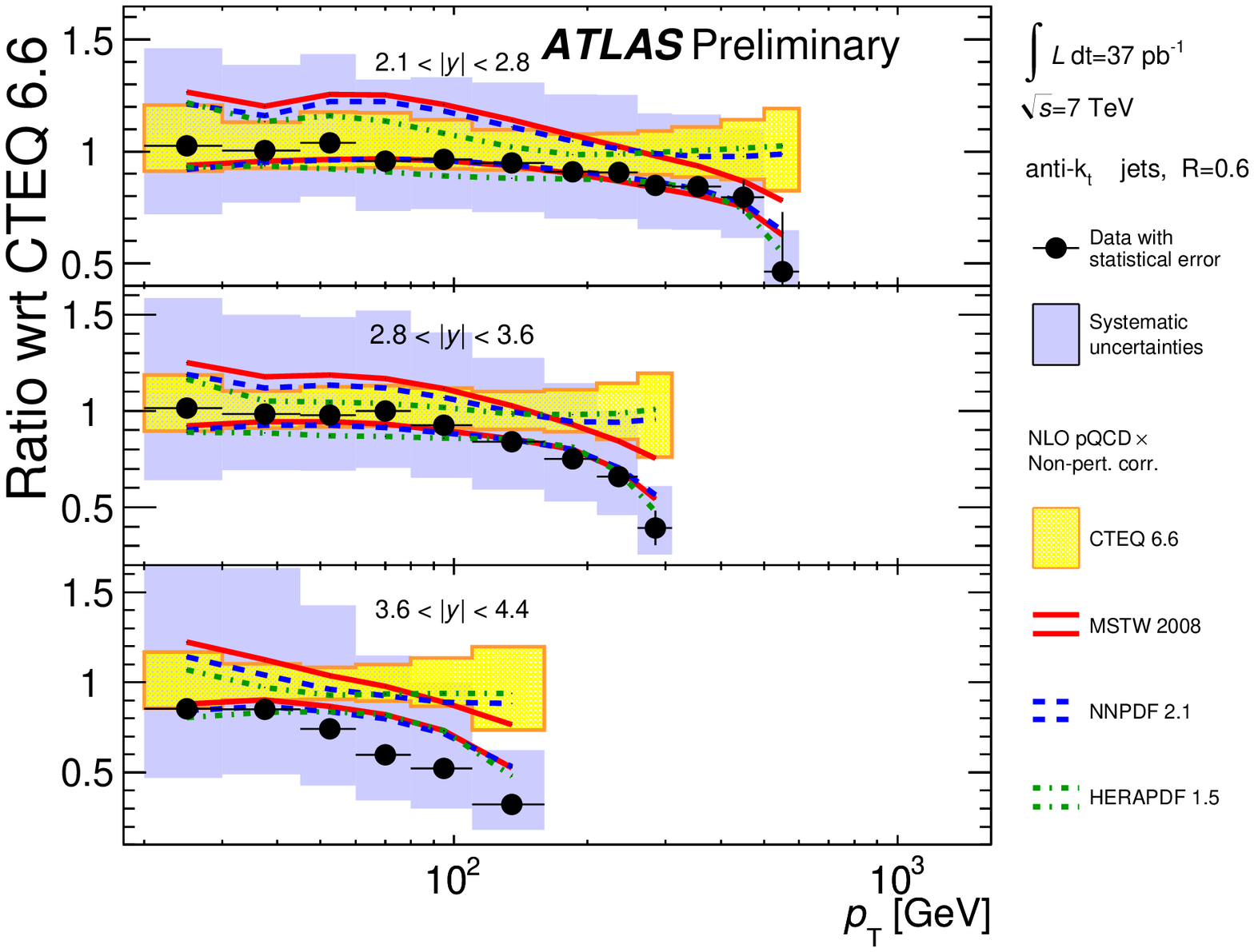}
\caption{Inclusive jet double-differential cross section as a function of jet $p_T$ in different regions of $|y|$ for jets identified using the anti-kt algorithm with R=0.6, measured at $\sqrt{s}=7$ TeV with ATLAS.  The theoretical error bands obtained using different PDF sets (CTEQ 6.6, MSTW 2008, NNPDF 2.1, HERAPDF 1.5) are shown.  The data points and the error bands are normalized to the theoretical predictions obtained by using the CTEQ  6.6 PDF set.}
\label{f:inclusiveATLAS}
\end{figure*}

Figure \ref{f:inclusiveATLAS} gives the comparison of ATLAS inclusive jet data 
with predictions from different parton distribution function (PDF) sets (CTEQ 6.6, MSTW 2008, NNPDF 2.1, HERAPDF 1.5).
  The data points and the error bands are normalized to the theoretical predictions obtained by using the CTEQ  6.6 PDF set.
Within the experimental uncertainties of 10-20\%, the data are found to be in 
agreement with
theoretical predictions from NLO pQCD.

\section{Dijets}
Events with two high $p_T$ jets (dijets) can be generated in proton-proton
collisions by parton-parton scattering.  They can be
characterized by the invariant
mass of the two jets, $M_{JJ}^2=x_1 \cdot x_2 \cdot s$, where $x_1,x_2$ are
the proton momentum fractions of the scattered partons. The dijet cross section
as a function of $M_{JJ}$ can be precisely calculated in pQCD, and can
be sensitive to new physics such as contact interactions or strongly-decaying
resonances.  

\begin{figure}
\includegraphics[width=80mm]{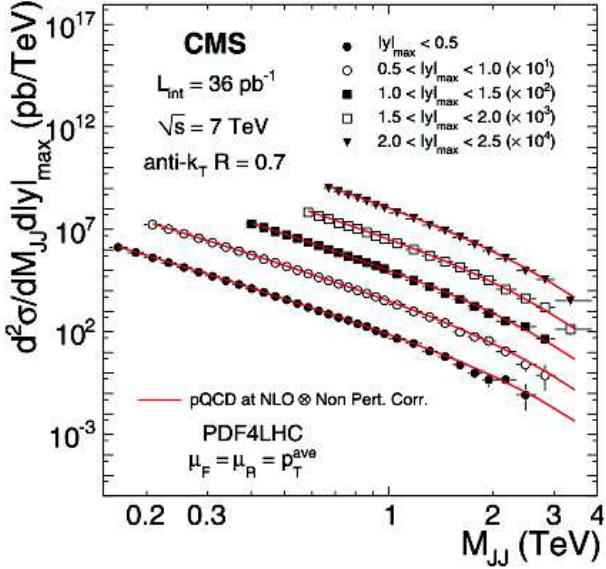}
\caption{Double-differential dijet production cross sections (points)
measured by CMS
as a function of the dijet invariant mass $M_{JJ}$ in bins of the variable $|y|_{max}$,
compared to the theoretical preductions (curves). The horizontal error bars
represent the bin widths, while the vertical error bars represent the
statistical uncertainties of the data.}
\label{f:CMSdijet}
\end{figure}

Figure \ref{f:CMSdijet} shows the double-differential dijet
production cross sections measured by CMS \cite{CMSdijets} as a function of $M_{JJ}$,
in bins of $|y|_{max}=max(|y_1|,|y_2|)$ of the two leading jets in the event.
Low values of $|y|_{max}$ tend to probe large-angle ($s$-channel) scattering, while
large values probe small-angle ($t$-channel) scattering.  
In this measurement, the dijet masses range from $M_{JJ}= 0.2$ to 3.5 TeV,
with corresponding parton momentum fractions
ranging from $8\cdot 10^{-4} \leq x_1 \cdot x_2 \leq 0.25$.
The NLO theoretical predictions,
at renormalisation and factorisation scales ($\mu_R$ and $\mu_F$) equal to
the average transverse momentum of the two jets ($p_T^{ave}€$),
together with non-perturbative corrections to take into account
hadronisation effects and multiparton interactions,
 are superimposed on the plots as curves.
The experimental systematic uncertainties are dominated by the JES uncertainty,
ranging from $3\%$ to $5\%$, giving an uncertainty on the cross section of
$15\%~(60\%)$ at $M_{JJ} = 0.2~(3)$ TeV.
The data and theoretical predictions
are found to be in good agreement within uncertainties.

\begin{figure}
\includegraphics[width=80mm]{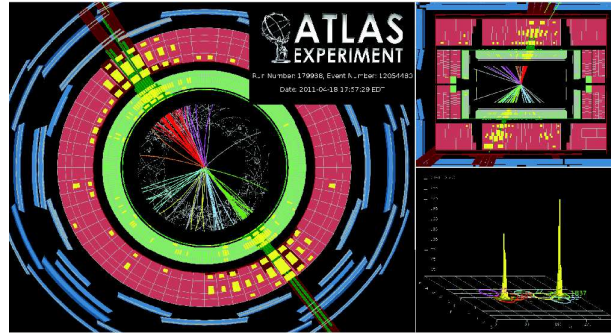}
\caption{A high-mass dijet event recorded by ATLAS in 2011.  The invariant mass of the dijet system is 4.0 TeV,
the two leading jets have ($p_T,y$) of (1.8 TeV, 0.3) and (1.8 TeV, -0.5), and the event $E_T^{miss}$= 100 GeV.}
\label{f:ATLASdijet}
\end{figure}

An event display of a high-mass dijet event recorded by ATLAS \cite{ATLASdijet} 
in 2011 is given
in Figure \ref{f:ATLASdijet}.  
The invariant mass of the dijet system is 4.0 TeV,
the two leading jets have ($p_T,y$) of (1.8 TeV, 0.3) and (1.8 TeV, -0.5), 
and the event has a missing transverse energy, $E_T^{miss}$, of 100 GeV.

\section{Jets in Association with Vector Bosons}
The production of massive vector bosons $V$ in association with jets
is a further test of pQCD.  Recently NLO predictions have become available \cite{Vplusjet} for $V+n$ jets,
where $n=3,4$ for $V=Z,W$ respectively.

\subsection{Jets + $W$ Bosons}
\begin{figure*}
\includegraphics[width=80mm]{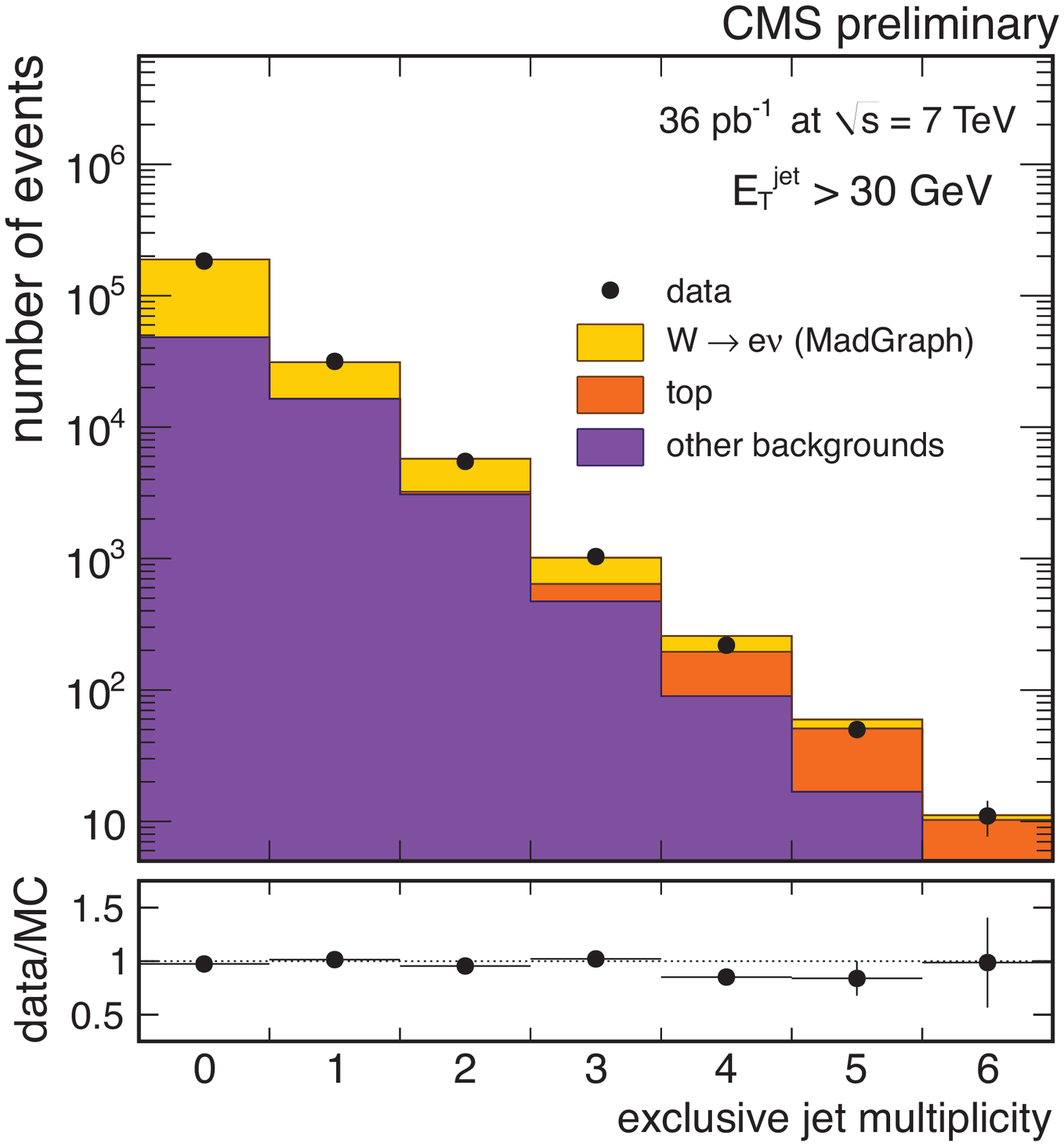}
\includegraphics[width=80mm]{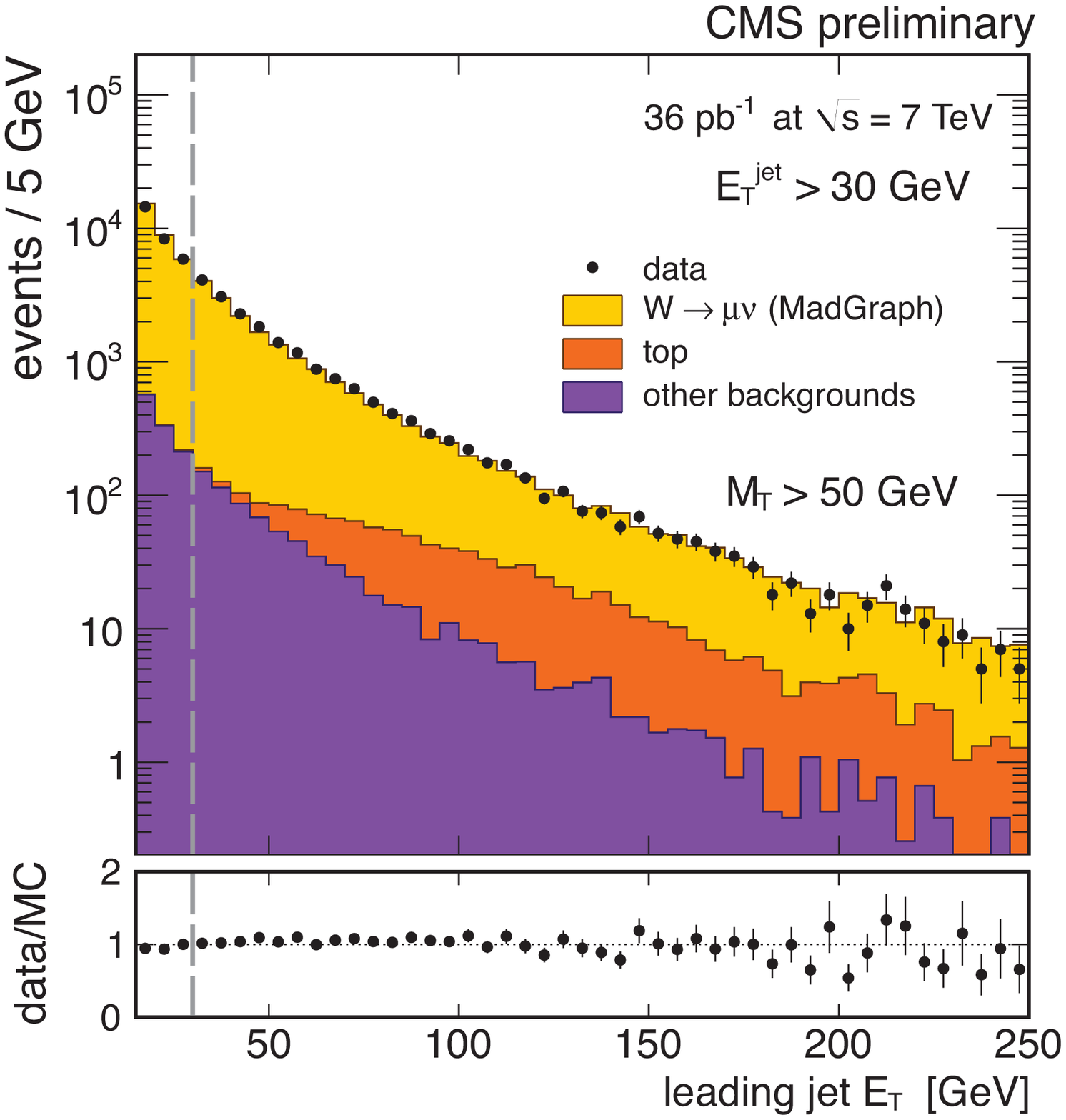}
\caption{The production of jets in association with $W$ bosons as measured by CMS.  
The left figure gives the raw $W$ event rate as a function of exclusive jet multiplicity
in the electron channel.   The right figure shows the 
uncorrected leading jet $E_T$
spectrum for the $W+1$ jet sample for the muon channel.
Data are plotted as points, and predictions from Monte Carlo
simulations for $W$ signal (MadGraph) and other backgrounds are shown 
as solid histograms.
The line at $p_T$= 30 GeV corresponds to the threshold
imposed for counting jets.}
\label{f:CMSWjet}
\end{figure*}

Figure \ref{f:CMSWjet} gives the measurements by CMS \cite{CMSWjet} of jets in association with
$W$ bosons.  The left figure shows the raw $W$ event rate in the electron channel
as a function of
 exclusive jet multiplicity, for jets with $E_T>$ 30 GeV.
   The right figure shows the leading jet transverse energy
 $E_T$
spectrum for the muon channel.
Data are plotted as points, and predictions from Monte Carlo
simulations for $W$ signal (MadGraph) and other backgrounds are shown.
In order to reduce backgrounds, a cut has been applied on the transverse mass
$M_T = \sqrt{2 p_T E_T^{miss}(1-cos{\Delta\phi})} > 50$ GeV, where
$\Delta\phi$ is the angle in the $xy$ plane between the lepton $p_T$ and
the $E_T^{miss}$ vector direction.
A good agreement is observed between data and theoretical predictions
for jet $E_T$ up to 250 GeV and final states with up to six jets.

\begin{figure*}
\includegraphics[width=80mm]{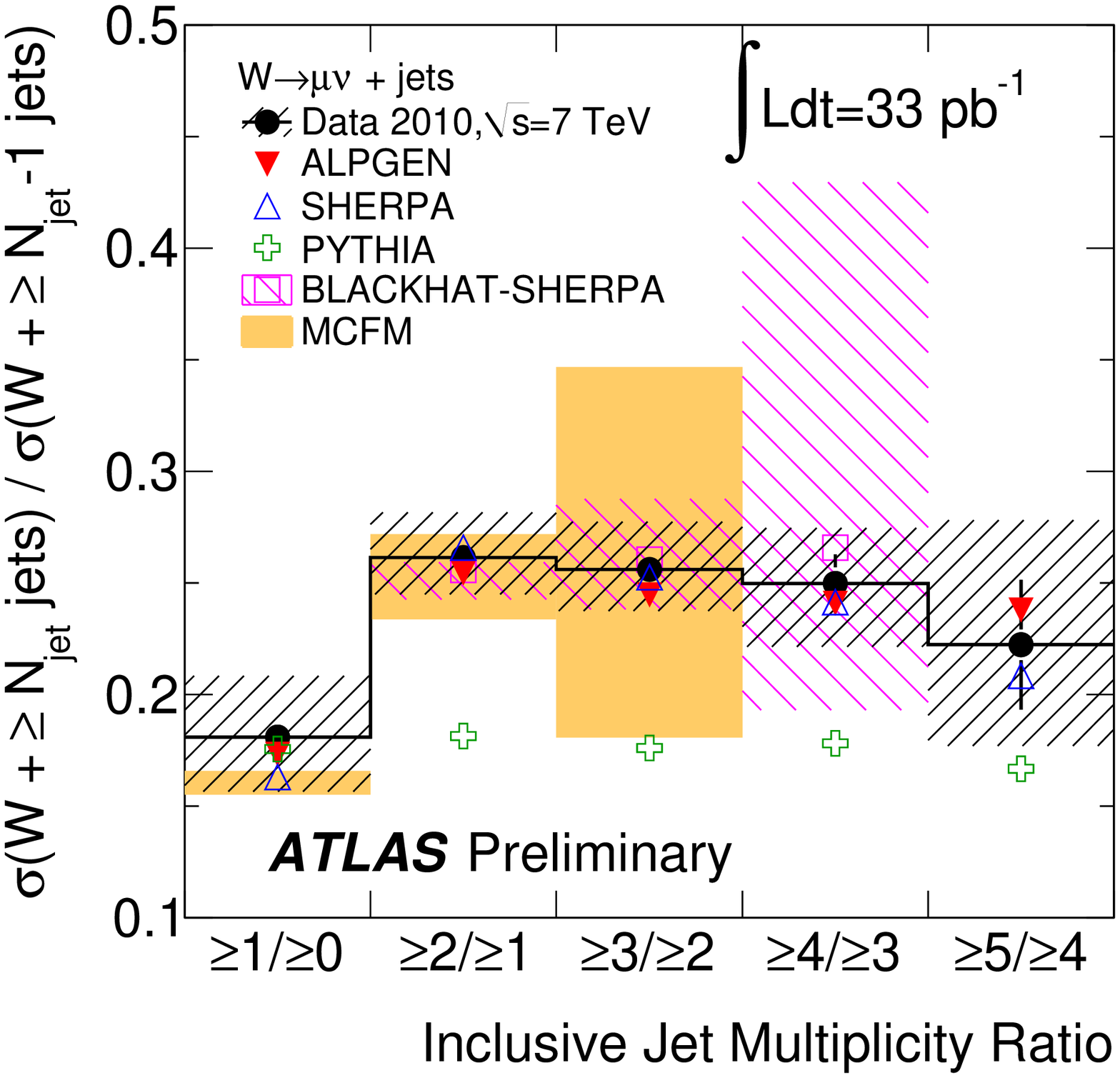}
\includegraphics[width=70mm]{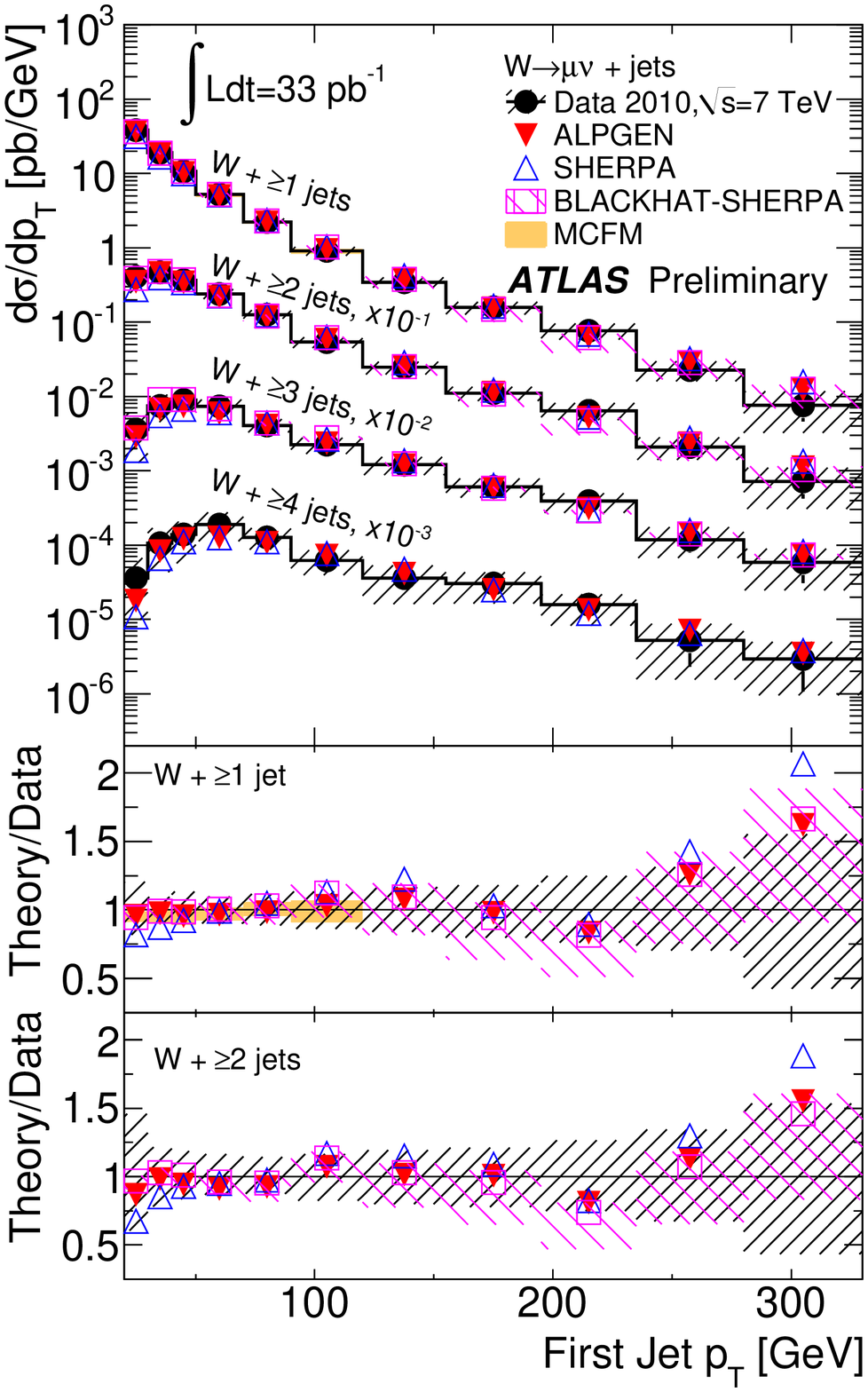}
\caption{The production of jets in association with $W$ bosons as measured by ATLAS.  
The left figure gives the $W$+jet cross section ratio as a function
of corrected jet multiplicity for the muon channel.
Data are plotted as points, and predictions from ALPGEN, SHERPA, PYTHIA,
MCFM, and BLACKHAT-SHERPA are superimposed as symbols.  Theoretical uncertainties
are shown only for MCFM and BLACKHAT-SHERPA.
The right figure gives the
$W$+jet cross section from ATLAS 
as a function of the $p_T$ of the first jet in the event 
for the muon channel. 
The $p_T$ of the first jet
is shown separately for events with $\geq 1$ jet to $\geq 4$ jets.}
\label{f:ATLASWjet}
\end{figure*}

Figure \ref{f:ATLASWjet} (left) shows the measurement by ATLAS \cite{ATLASWjet}
of jets in association with a $W$ boson expressed
as a ratio of cross sections $\sigma(W+\geq N_{jet}) / \sigma (W+\geq N_{jet}-1)$
for inclusive jet multiplicities $N_{jet}=1-5$.
The figure gives the $W$+jet cross section ratio as a function
of corrected jet multiplicity for the muon channel.
Figure \ref{f:ATLASWjet} (right) gives the
$W$+jet cross section as a function of the $p_T$ of the first jet in the event 
for the muon channel.  The $p_T$ of the first jet
is shown separately for events with $\geq 1$ jet to $\geq 4$ jets.
Data are plotted as points, and predictions from ALPGEN, SHERPA, PYTHIA,
MCFM, and BLACKHAT-SHERPA are superimposed as symbols.  Theoretical uncertainties
are shown only for MCFM and BLACKHAT-SHERPA.
As PYTHIA is a LO calculation, it does not provide a good description
of the data for jet multiplicities greater than one.
However, the NLO predictions from MCFM and BLACKHAT-SHERPA, and
multi-parton matrix element generators ALPGEN and SHERPA,
are found to be in good agreement with the data.

\subsection{Ratio of Cross Sections for Jets + $W/Z$ Bosons}

\begin{figure}
\includegraphics[width=90mm]{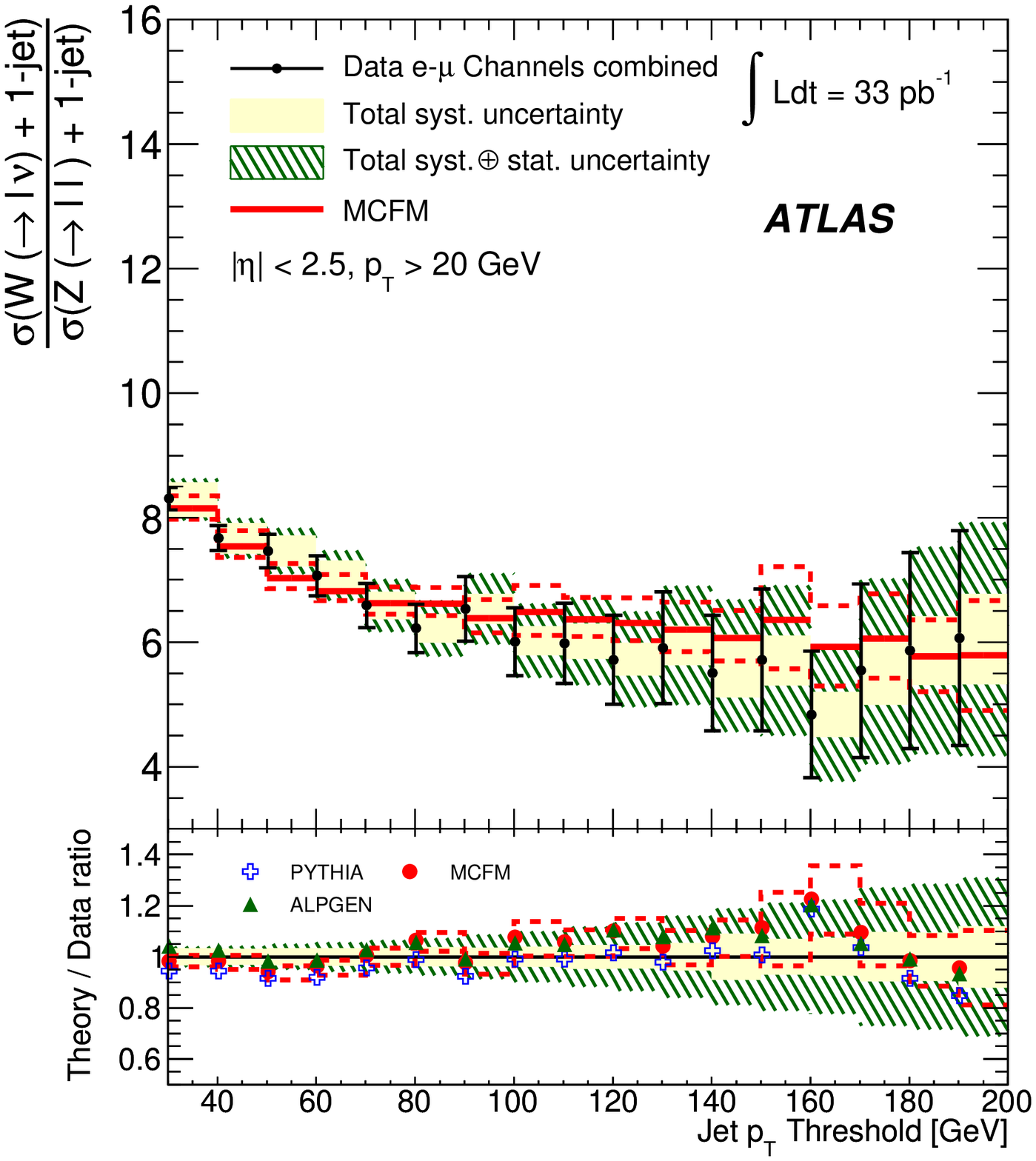}
\caption{The ratio of cross sections measured by ATLAS
for events with one jet and a $W$ boson
in the final state, compared to those with one jet and a $Z$ boson in the final state,
as a function of the jet $p_T$ threshold.
Data are shown with black error bars indicating the statistical uncertainties.
The yellow band shows all systematic uncertainties added in quadrature
and the green bank shows statistical and systematic uncertainties added in quadrature.
The theory uncertainty (dashed line) includes contributions from PDF
and renormalisation and factorisation scales.
}
\label{f:ATLASRatioWZjet}
\end{figure}

Figure \ref{f:ATLASRatioWZjet} shows the ratio of cross sections measured by ATLAS
\cite{ATLASRatioWZjet}
for events with one jet and a $W$ boson
in the final state compared to those with one jet and a $Z$ boson in the final state,
as a function of the jet $p_T$ threshold.
This ratio is useful as a number of theoretical and experimental 
uncertainties cancel.
As the jet $p_T$ threshold increases, the ratio is expected
to decrease, as the effective scale of the interaction
becomes large compared to the difference in boson masses;
this dependence is observed in the data.
The electron and muon channel results were found
to be compatible, and are combined in a common
fiducial region of $|\eta_{lepton}| < 2.5$, $p_T^{lepton}>20$ GeV.
The results are
compared to predictions from MCFM (corrected to particle level),
and found to be in agreement.

\subsection{$W+b$ Jets}

\begin{figure*}
\includegraphics[width=80mm]{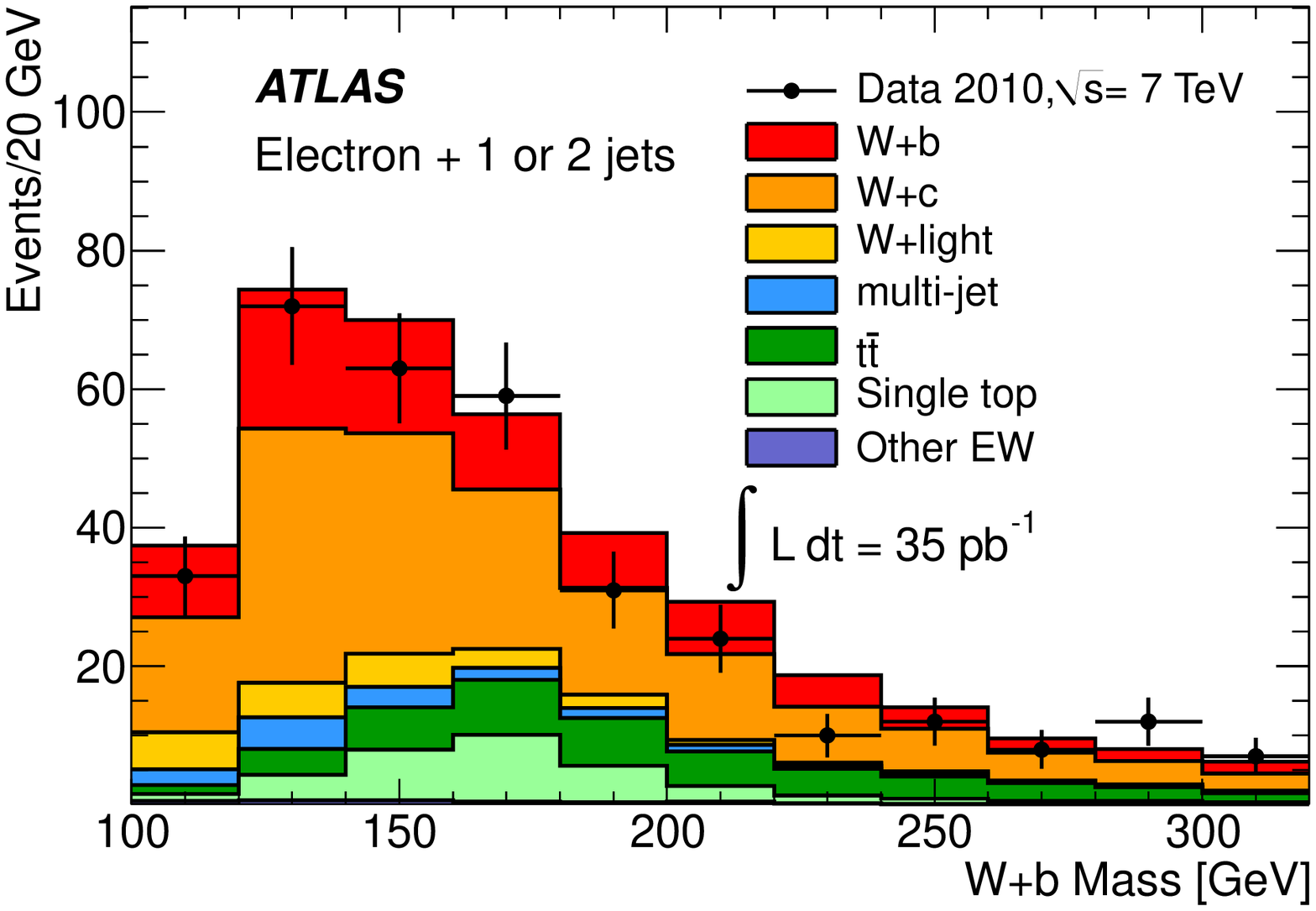}
\includegraphics[width=80mm]{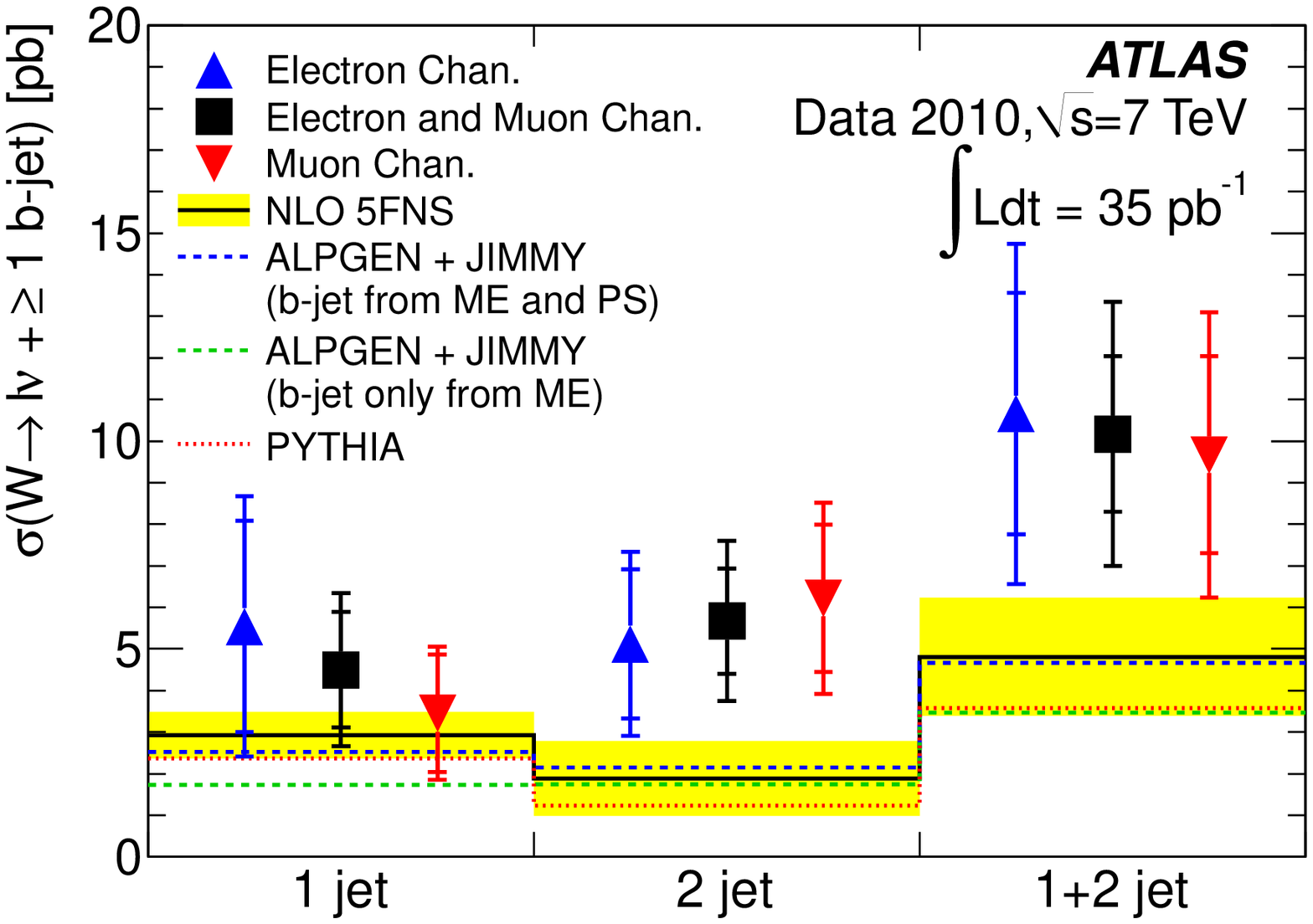}
\caption{Production of $W$ bosons in association with a $b$-jet
measured by ATLAS.
The left figure gives the invariant mass of the $W+b$-jet system
in the electron channel. 
The right plot gives the measured fiducial cross section
in the 1, 2, and 1+2 exclusive bins.  The measurements
are compared with NLO predictions; the yellow band represents the total
uncertainty on the prediction.  LO predictions from ALPGEN+HERWIG+JIMMY
and PYTHIA are given as well.
}
\label{f:ATLASWbjet}
\end{figure*}

$W+b$-jet production is a large background to searches for the Higgs boson
in $WH$ production with the decay $H \rightarrow b \overline{b}$.
Previous measurements in proton-antiproton collisions by CDF \cite{CDFWbjet}
indicate a larger measured cross section than the NLO QCD prediction.
The measurement by ATLAS \cite{ATLASWbjet} 
of the cross section for the production of a $W$ boson
with one or two jets is shown in Figure \ref{f:ATLASWbjet}.
The left figure gives the invariant mass of the $W+b$-jet system
in the electron channel.  The right figure shows the measured
fiducial cross section in the 1, 2, and 1+2 jet exclusive bins. 
Anti-kt jets with $R=0.4$ are reconstructed,
with $p_T>$25 GeV and $|y|<2.1$ (see Table 1 of \cite{ATLASWbjet} for the 
full definition of the fiducial region).
The measurements are compared with NLO predictions,
with systematic uncertainties from the renormalisation scale,
factorisation scale, PDF set, and non-perturbative corrections
combined in quadrature.
The results are consistent with NLO predictions at the 1.5 $\sigma$ level.

\subsection{$Z+b$ Jets}
\begin{figure*}
\includegraphics[width=40mm]{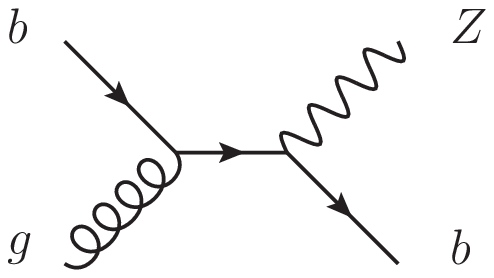}
\includegraphics[width=40mm]{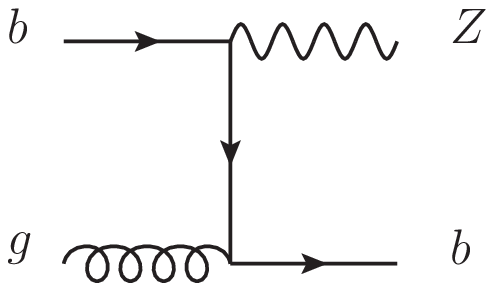}
\includegraphics[width=40mm]{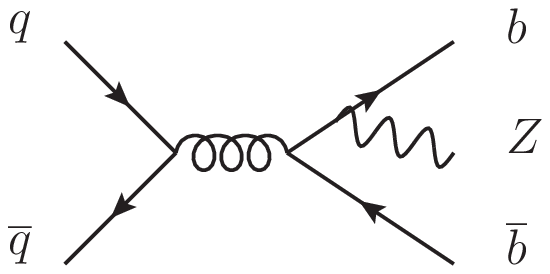}
\includegraphics[width=40mm]{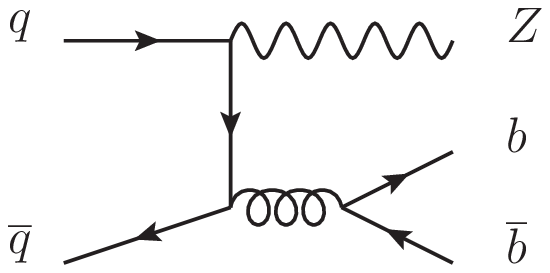}
\caption{Main diagrams for associated production at the LHC
of a $Z$ boson and one or more
$b$-jets.
}
\label{f:ZbFeyn}
\end{figure*}

The production of one or more $b$-jets in association with a $Z$ boson
is another significant background to searches for new physics at the LHC.
Figure \ref{f:ZbFeyn} shows the main diagrams that contribute to $Z+b$ production
at the LHC.  The first 2 diagrams have an initial state $b$-quark, while the last
2 have a $b\overline{b}$ pair produced explicitly in the final state.
A summary of measurements \cite{Zplusbjets} from the LHC and Tevatron of the average number of 
$b-$jets produced in association with a $Z$ boson is
given in Table \ref{t:Zbjet}, compared to theoretical predictions from MCFM.
In the case of ATLAS, the ratio is given 
with respect to the total $Z$ production cross section,
while for CMS, CDF, and D0 it is given with respect to the $Z+$jet production
cross section.  The MCFM NLO prediction is in agreement with the data.

\begin{figure*}
\includegraphics[width=130mm]{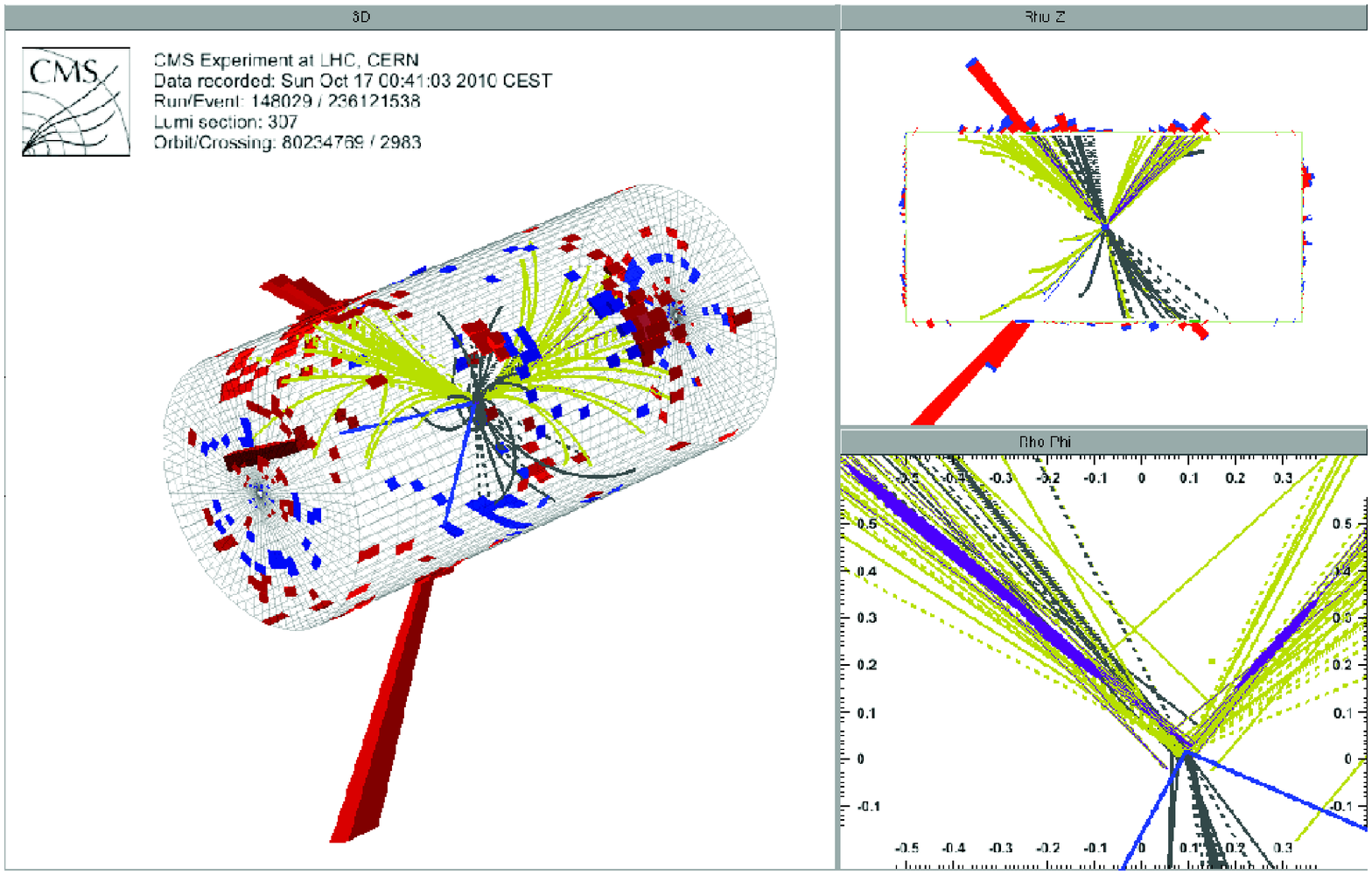}
\caption{An event display of a $Z\rightarrow e^+ e^- + b-$jet candidate recorded
by CMS.  The bottom right view shows the 2 $b$-tagged jets as dark enlongated ovals.
}
\label{f:CMSZbjet}
\end{figure*}

Finally, an example
event display of a $Z\rightarrow e^+ e^- + b-$jet candidate recorded
by CMS is shown in Figure \ref{f:CMSZbjet}.
The jets have $p_T$ of 121, 62, 42, and 36 GeV, and the leptons have
$p_T$ of 119 and 32 GeV.  The dilepton invariant 
mass is 90.3 GeV, the mass of the $b\overline{b}$ system is 192 GeV, and the 
mass of the $Zb\overline{b}$ system is 400 GeV. 

\begin{table*}[b]
\begin{center}
\caption{\label{Zplusbjets}Measured ratios of $Z$ plus $b$-jets  versus overall $Z$ production, compared to theoretical predictions (MCFM).}
\begin{tabular}{|l|c|c|c|}
\hline \textbf{Collaboration} & \textbf{Ratio} & \textbf{Data} &
\textbf{MCFM}
\\
\hline ATLAS & $\sigma (Z+b) / \sigma (Z)$ & $(7.6 \pm 0.18 \pm 0.15) \times 10^{-3}$ & $(8.81 \pm 1.1) \times 10^{-3}$  \\
\hline CMS & $\sigma (Z+b) / \sigma (Z+j)$ & $5.4 \pm 0.16 \% $ & $4.3 \pm 0.16 \% $  \\
\hline CDF & $\sigma (Z+b) / \sigma (Z+j)$ & $2.24 \pm 0.16 \pm 0.27 \% $ & $2.2 \% $  \\
\hline D0 & $\sigma (Z+b) / \sigma (Z+j)$ & $1.93 \pm 0.22 \pm 0.15 \% $ & $1.92 \pm 0.22 \% $  \\
\hline
\end{tabular}
\label{t:Zbjet}
\end{center}
\end{table*}

\section{Conclusion}
Experiments at the LHC and Tevatron have performed
an exhaustive set of measurements of jet production,
of which a small sample are documented here,
including inclusive jet, dijets,
jets in association with $W$ and $Z$ bosons,
and $b$-jets.  These measurements provide extensive
probes of perturbative QCD, which to date
has passed these tests with flying colours.

\bigskip 
\begin{acknowledgments}
The author would like to thank the people from ATLAS, CMS, CDF, and D0
who have made these measurements possible, and the
organizers of the Physics in Collision Symposium.
\end{acknowledgments}

\bigskip 
\bibliography{basename of .bib file}

\end{document}